\documentclass[onecolumn]{article}

\usepackage{graphicx}
\usepackage{ amssymb}
\usepackage[ansinew]{inputenc}
\usepackage{authblk}
\RequirePackage{subfig}

\begin{document}

\title{Relaxation of axially confined 400 GeV/c protons to planar channeling in a bent crystal}

\author[1]{L. Bandiera} 
\author[1]{A. Mazzolari} 
\author[1]{E. Bagli} 
\author[1]{G. Germogli} 
\author[1]{V. Guidi} 
\author[1,6]{A. Sytov} 
\author[2]{I. V. Kirillin} 
\author[2]{N. F. Shul'ga} 
\author[3]{A. Berra} 
\author[3]{D. Lietti} 
\author[3]{M. Prest} 
\author[4]{D. De Salvador} 
\author[5]{E. Vallazza}

\affil[1]{INFN Sezione di Ferrara, Dipartimento di Fisica e Scienze della Terra, Universit{\`a} di Ferrara Via Saragat 1, 44122 Ferrara, Italy}
\affil[2]{Akhiezer Institute for Theoretical Physics, National Science Center ``Kharkov Institute of Physics and Technology'', Akademicheskaya Str.,~1, 61108 Kharkov, Ukraine $\&$ V.N. Karazin Kharkov National University, Svobody Sq. 4, 61022, Kharkov, Ukraine}
\affil[3]{Universit{\`a} dell'Insubria, via Valleggio 11, 22100 Como, Italy $\&$ INFN Sezione di Milano Bicocca, Piazza della Scienza 3, 20126 Milano, Italy }
\affil[4]{INFN Laboratori Nazionali di Legnaro, Viale dell'Universit{\`a} 2, 35020 Legnaro, Italy $\&$ Dipartimento di Fisica, Universit{\`a} di Padova, Via Marzolo 8, 35131 Padova, Italy}
\affil[5]{INFN Sezione di Trieste, Via Valerio 2, 34127 Trieste, Italy }
\affil[6]{Research Institute for Nuclear Problems, Belarusian State University, Bobruiskaya street, 11, Minsk 220030, Belarus}

\maketitle

\begin{abstract}
An investigation on the mechanism of relaxation of axially confined 400 GeV/c protons to planar channeling in a bent crystal was carried out at the extracted line H8 from CERN Super Proton Synchrotron. The experimental results were critically compared to computer simulations, showing a good agreement. We firmly individuated a necessary condition for the exploitation of axial confinement or its relaxation for particle beam manipulation in high-energy accelerators. We demonstrated that with a short bent crystal, aligned with one of its main axis to the beam direction, it is possible to realize either a total beam steerer or a beam splitter with adjustable intensity. In particular, in the latter case, a complete relaxation from axial confinement to planar channeling takes place, resulting in beam splitting into the two strongest skew planar channels. 
\end{abstract}

\section{Introduction}

A charged particle moving inside a crystal nearly aligned with crystallographic directions or planes may suffer correlated collisions with neighboring lattice atoms. As a result, its dynamics can be described by the continuous potential of atomic strings/planes within which a charged particle trajectory may be trapped, leading to the process of  {\it channeling}. Channeling may occur as the particle incidence angle with the crystal axes (planes) is lower than the critical angle introduced by Lindhard, $\psi_c=\sqrt{{2U_{0,ax}}/{pv}}$ ($\theta_c=\sqrt{{2U_{0,pl}}/{pv}}$), $U_0$ being the continuous potential well depth, $p$ and $v$ the particle momentum and velocity, respectively \cite{Dansk.Fys.34.14}. Since the the axial potential well $U_{0,ax}$ is deeper than the planar $U_{0,pl}$, $\psi_c$ is larger than $\theta_c$.

After the pioneering work of Tsyganov in 1979 \cite{Tsyganov682,Tsyganov684}, the usage of {\it bent} crystals has been deeply investigated for manipulation of the trajectories of charged beams in high-energy particle accelerators. Indeed, it is possible to effectively deflect the particle beam direction by a small-sized bent crystal, in place of a cumbersome and expensive super-conductive magnet \cite{Biryukov,Biryukov200523,Scandale:1357606}. 

Various mechanisms of deflection of charged particles by a bent crystal have been investigated over the years. These mechanisms involve either bound and unbound particle states in both the axial and planar potential wells. The deflection mechanisms involving bound states are axial (AC) and planar (PC) channeling, while over-barrier particles may be deflected by the stochastic mechanism of deflection (SD)\cite{Shul'ga1995373} caused by multiple scattering with atomic strings (the so-called \textit{doughnut} scattering) and by volume reflection (VR) from bent crystal planes \cite{Taratin1987425}. In view of applications to beam steering, most of the studies have been focused on the planar cases due to the easier experimental requirements for their implementation \cite{scandale2010first}. Nevertheless, the exploitation of the axial mechanism of deflection may bring several advantages. Indeed, the stronger potential of the axes leads to a larger angular acceptance and a higher deflection efficiency than for PC. In particular, in case of axial alignment, most of the particles are deflected through SD rather than by AC. Since SD consists of the deflection of unbound particles, the mechanism of dechanneling, i.e., the kick out of channeled (bound) particles from the potential well, does not appreciably affect the process of particles steering as it does for PC. Furthermore, if one compares SD with VR, it can be noticed that even if the latter has a larger acceptance, the achievable deflection angle under stochastic regime can be far larger than for VR, for which is of the order of $\theta_c$. 

SD was predicted in \cite{grinenko1991turning} and allows deflecting the whole beam if the crystal is bent with an angle $\alpha$, which obeys the following inequality \cite{shulga-nuovo}
\begin{equation}
 \alpha < \alpha_{st}=\frac{2R\psi^2_c}{l_0},
 \label{eq:max}
\end{equation}
where $\alpha_{st}$ is the maximum angle achievable through SD, $R$ is the radius of crystal curvature, $l_0=4/(\pi^2ndR_a\psi_c)$, $n$ being the concentration of atoms in the crystal, $d$ the distance between neighboring atoms in the atomic string and $R_a$ the atomic screening radius. Condition (1) was found without taking into account incoherent scattering, i.e., $\alpha_{st}$ is a function of particle energy and $R$ and \textit{not} of the crystal length. 

\begin{figure}[ht!]
\centering
\includegraphics[width=1\columnwidth]{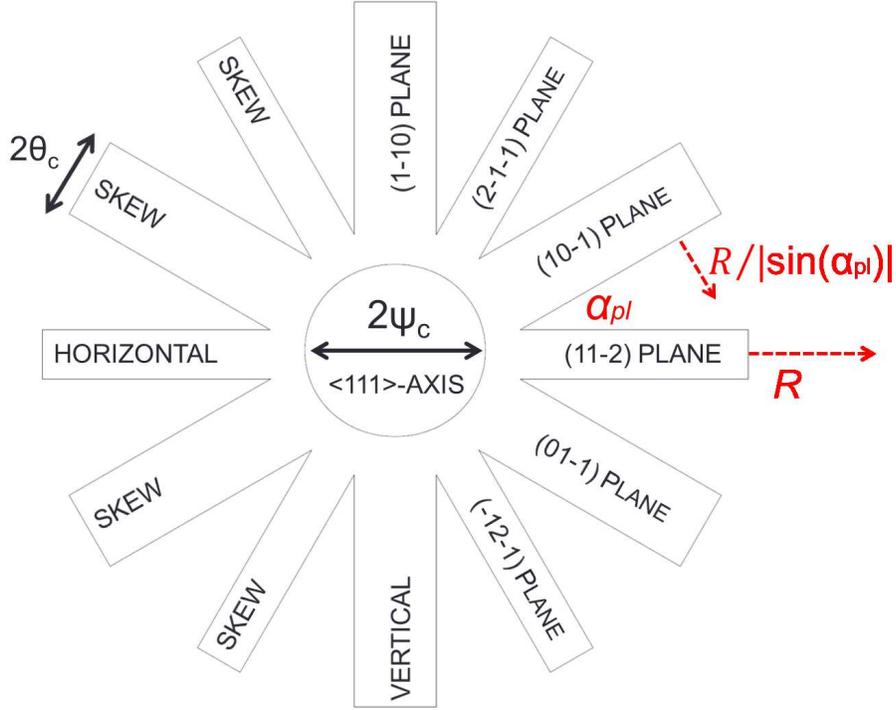}\\
\caption{Angular region around $\left< 111 \right>$ axis, where the strongest (110) and (211) planes are underlined. $R$ is the radius of curvature of both the $\left< 111 \right>$ axis and the ($1\bar{1}0$) vertical plane, while $R/\sin(|\alpha_{pl}|)$ represents the bending radius of each skew plane, $\alpha_{pl}$ being the angle between the skew plane and the ($11\bar{2}$) horizontal plane.}\label{skew}
\end{figure}

Pioneering experiments on axial channeling \cite{Bak19841,baurichter1996new} failed to observe an efficient beam deflection at the nominal bending angle because the Eq. (\ref{eq:max}) was not fulfilled, i.e., $\alpha \gg \alpha_{st}$. Indeed, the experiments involving positive particles highlighted strong feed-in of particles into the channels of \textit{skew planes} (e.g., see Fig. 10 in Ref. \cite{baurichter1996new}). In order to visualize the effect of skew planes, Fig. \ref{skew} represents the strongest (110) and (211) planes intercepting the $\left< 111 \right>$ axis (the same used in the experiments described below in the text). Particles deflection occurs along the flat horizontal ($11\bar{2}$) plane. All other planes result to be bent with a curvature radius $R/|\sin(\alpha_{pl})|$, $\alpha_{pl}$ being the inclination angle of the plane with respect to the horizontal one. The vertical plane ($1\bar{1}0$) is bent with radius $R$ as well as the $\left< 111 \right>$ axis.

Only recently, SD was experimentally observed for either positively \cite{PhysRevLett.101.164801} or negatively \cite{Scandale2009301} charged particles, owing to the advent of a new generation of bent crystals \cite{AfoninU70JETP,guidi:113534}, short enough to efficiently deflect high-energy particles up to the nominal bending angle of the crystal. 

In particular, in Ref.\cite{PhysRevLett.101.164801} the potential of axial confinement for steering of positive beams was highlighted, despite that a strong feed in of particles into skew planar channels was recorded. Indeed, a total 90$\%$ of deflection efficiency was reached due to SD accompanied by relaxation into skew planes. Such a high value has never been achieved in the case of PC. Therefore, SD accompanied by the relaxation to skew planes can be a valuable alternative for beam steering in accelerators.
 
Apart from the result presented in Ref. \cite{PhysRevLett.101.164801,Scandale2009301}, no other data collected with short crystals suitable for applications are available in the literature. As pointed out in \cite{Carrigan25S1}, there is a need for deeper understanding about axial channeling for its exploitation in beam manipulation.

In this paper we present an investigation through experiments and simulations on the axial-to-planar channeling relaxation of high-energy positive particles in bent crystals and its dependence on crystal curvature. With this, we also addressed possible schemes for manipulation of particle beams, such as steering or splitting. 

\section{Experiment with 400 GeV/c protons}

An experiment was carried out at the external beamline H8 of CERN-Super Proton Synchrotron, where a primary beam of 400 GeV/c protons is available. Two $1$x$55$x$2$ mm$^3$ strip-like Si crystals ($L = 2mm$ thick along the beam direction) with the largest faces oriented parallel to the ($1\bar{1}0$) planes were used. The crystals were fabricated according to the procedure described in Ref. \cite{baricordi:061908,0022-3727-41-24-245501} and bent through anticlastic deformation \cite{guidi:113534} to curvature radii $R_1$ = ($30.30\pm0.05$) m and $R_2$ = ($6.90\pm0.05$) m. Given $\psi_c \approx $ 21 $\mu$rad for the $\left< 111 \right>$ axes, from condition (1) it comes out that $\alpha_{st,1} \approx 604$ $\mu$rad and $\alpha_{st,2} \approx$ 58$\mu$rad for the two crystals, respectively. Since the bending angles are equal to $\alpha_1 = L/R_1 =$ 66 $\mu$rad and $\alpha_2 = 290$ $\mu$rad, it follows that the first crystal fulfills condition (1) for SD, while the second does not.

The two crystals were mounted on a high-precision goniometer with the possibility to be aligned in either horizontal or vertical direction. Deflection was measured by a telescope system based on Si double-sided microstrip detectors \cite{Lietti2013527} with an angular resolution of 4 $\mu$rad and 4.7 $\mu$rad in the horizontal and vertical deflection angle, respectively. The experimental setup was based on the \textit{part A} of the setup in \cite{Bandiera2013135}. By exploiting the horizontal rotational movement of the goniometer, we first attained PC in the $(1\bar{1}0)$ planes, then by scanning the vertical rotational movement, the crystals were aligned with the $\left< 111 \right>$ axis.  During the offline analysis, strip torsion induced by holder mechanical imperfections \cite{bagli2010fabrication} was evaluated and taken into account. After that, an angular cut of $\pm$ 5 $\mu$rad in either $x$ or $y$ direction of the incident beam was set in order to select only the particles aligned with the $\left< 111 \right>$ crystal axes.

\begin{figure}[ht!]
\centering
\includegraphics[width=1\columnwidth]{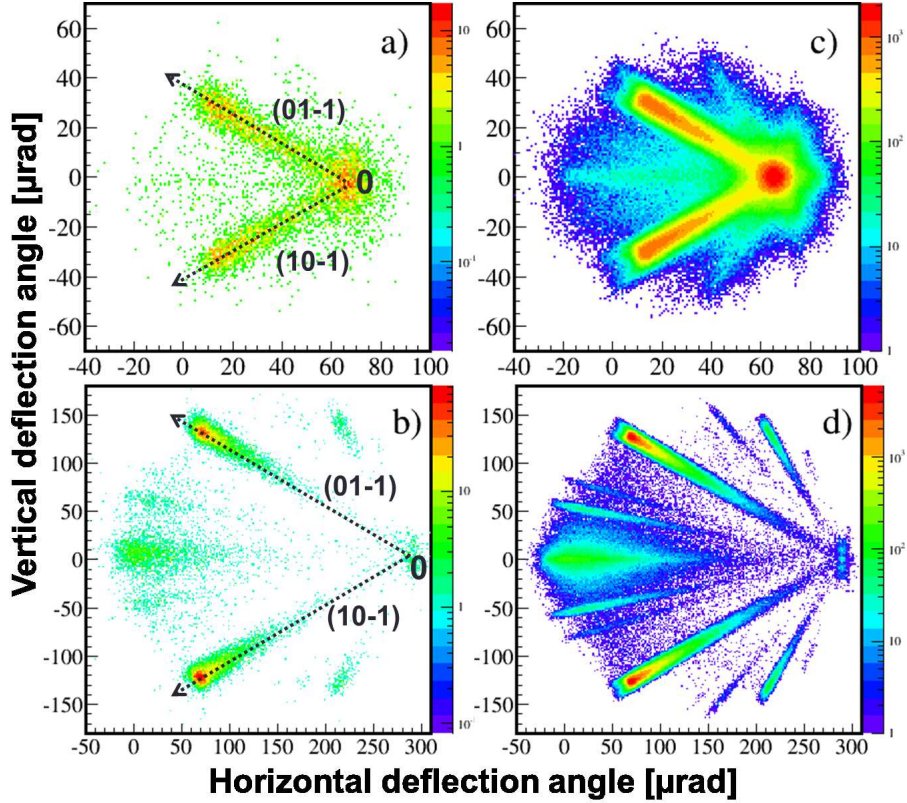}\\
\caption{(a)-(b) Experimental angular distribution of 400 GeV/c protons after interaction with the 2 mm long Si crystals, aligned with the $\left< 111 \right>$ axis, bent with curvature radii equal to $R_1$ = 30.3 m (a) and $R_2$ = 6.9 m (b), respectively. (c)-(d) Simulated angular distribution for the (a)-(b) cases. The dotted arrows in (a)-(b) highlight the directions of deflection for the $(01\bar{1})$ and the $(10\bar{1})$ skew planes. (Colors online show the intensity of the deflected beam in logarithmic scale.)}\label{fig:2D-dist}
\end{figure}

Figs. \ref{fig:2D-dist}a-b show the experimental angular distributions of 400 GeV/c protons after interaction with the crystals bent to $R_1$ (a) and $R_2$ (b), respectively. The $x$-axis lies in the $(11\bar{2})$ plane, while the $y$-axis in the $(1\bar{1}0)$ plane. In Fig. \ref{fig:2D-dist}a, a large part of the protons traversing the crystal kept under SD regime (about 30 $\%$) and were deflected to the nominal bending angle $\alpha_1$, while other protons relaxed to PC during their motion through the crystal. On the contrary, Fig. \ref{fig:2D-dist}b highlights that the overwhelming majority of the protons relaxed to planar channels $(01\bar{1})$ and $(10\bar{1})$. The difference between the distributions in plots (a) and (b) owes to the fulfillment (for $R_1$) or not-fulfillment (for $R_2$) of condition (\ref{eq:max}). In both cases, almost no particle remains with the direction of the incoming beam and about a fraction of 98$\%$ of the beam is deflected with an horizontal angle $>$ 0, i.e., a \textit{total steering} of the beam was accomplished. In case (b), most of the particles (80$\%$) were captured by the two strongest (110) skew planes and thus well separated at the crystal exit by an angle equal to 250 $\mu$rad, resulting in efficient beam \textit{splitting} other than \textit{steering}.

An intriguing feature one can observe in Figs. \ref{fig:2D-dist}a-b is the lack of dechanneling from skew planes. Indeed, during the relaxation process, protons are captured in a skew planar channel without approaching close to atomic strings, thereby entering close to the minimum of the planar potential well. In other words, the lack of dechanneling, which is a feature of SD, is preserved even when the particles relax to skew planes. Such particles may be dechanneled later from skew planes due to scattering with valence electrons. The usage of a crystal much shorter (2 mm) than the \textit{electronic dechanneling} length ($l_D \simeq 220$ mm for 400 GeV/c protons in the field of (110) planes \cite{Biryukov}) prevented the loss of particles from skew planes. 

\section{Theoretical investigation and discussion}

In order to investigate more deeply the features underlined in the above section, a Monte Carlo simulation has been worked out, by using the same code as in Ref. \cite{shulga2011dynamical}. The code solves the equation of motion in the field of continuum potential through numerical integration and also takes into account the contribution of incoherent scattering with atomic nuclei and electrons. Figs. \ref{fig:2D-dist}c-d display the simulated deflection distributions for both crystals, taking into account the experimental resolution. 
The comparison between experimental and simulated results in Fig. \ref{fig:2D-dist} exhibits a very good agreement. The largest statistics achieved through simulations also highlights the escape of protons toward a wealth of minor skew planes other than the (110)s, especially for the case of the most bent crystal (see Fig. \ref{fig:2D-dist}d).


\begin{figure} 
\centering
 \includegraphics[width=1\columnwidth]{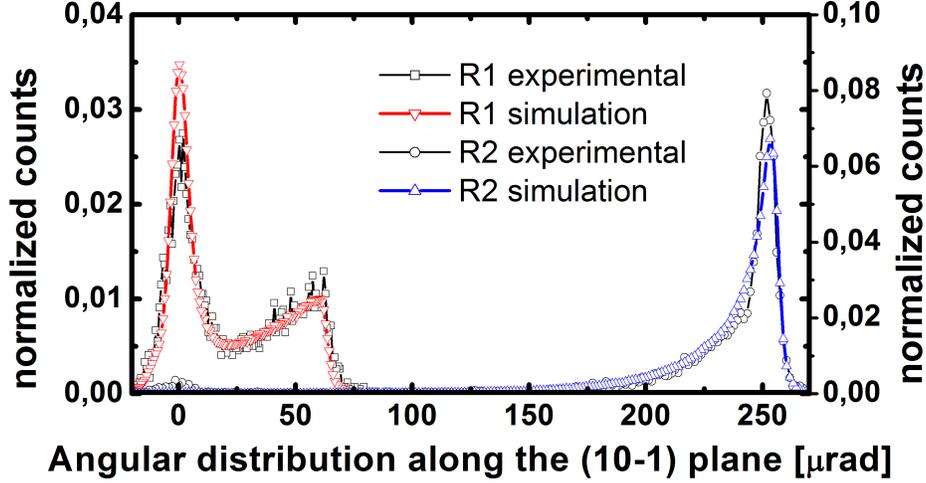}\\
  \caption {Experimental and simulated angular distributions of particles in the skew planar channels $(10\bar{1})$, respectively. Left Y-scale: squares and down-triangles are for experimental and simulated distribution for $R_1$= 30.3 m. Right Y-scale: circles and up-triangles are the same for $R_2$= 6.9 m. }\label{fig:skew}
\end{figure}

A deeper insight into the mechanism of axial-to-planar relaxation has been carried out by analyzing the particle distributions in Fig. \ref{fig:2D-dist} along one of the strongest skew plane, i.e., the $(10\bar{1})$ plane. The results are displayed in Fig. \ref{fig:skew} for both experiments and simulations. The origin of the coordinates corresponds to the direction of $\left< 111 \right>$ axes at the end of the crystal, i.e., the origin of the two dashed arrows in Figs. \ref{fig:2D-dist}a-b. In both cases in Fig. \ref{fig:skew}, the regions between the final axis positions (zero) and the farthest points from the axes along the $(10\bar{1})$ planes are populated by the particles that have undergone relaxation from SD to PC in their motion through the crystal. It is clear that the escape velocity from SD to skew planes increases while $R$ decreases. By assuming that the rate of particle escaping from stochastic deflection is proportional by a factor $-C$ to the number of particles that are in this regime, $N$, one can write the following equation
\begin{equation}
\frac{d N}{d l} = - C N,
\label{eq:velocity}
\end{equation}
where $l$ is the length of the particle path inside the crystal.
By solving Eq. \ref{eq:velocity} one obtains:
\begin{equation}
N (l) = N_0 e^{- C l} = N_0 e^{- l / l_R},
\label{eq:number-SD}
\end{equation}
where $l_R$ is the length within which the number of particles in SD regime $e$-folds, i.e., the \textit{relaxation length}, and $N_0$ the total number of particles captured into the SD regime at the crystal entrance (all the particles in the cases under consideration here).
The number of particles captured under channeling regime in all the skew planes, $N_{pl}$, vs. $l$ is derived from Eq. \ref{eq:number-SD} as follows:
\begin{equation}
N_{pl} (l) = N_0 \left( 1 - e^{- l / l_R} \right).
\label{eq:number-PC}
\end{equation}
This relationship determines the exponential form between the two peaks in the distributions of Fig. \ref{fig:skew}.

For a crystal bent to a curvature radius $R$, the relaxation length $l_R$ determines the maximum crystal length for efficient steering of particles at the full bending angle $\alpha = L/R$. $l_R$ is a more direct physical quantity than $\alpha_{st}$ to design an optimal crystal for beam manipulation under SD. Moreover, differently from $\alpha_{st}$, this quantity accounts for the contribution of incoherent scattering with atomic nuclei and electrons. One can replace the ideal condition (1) for SD with the more useful relation $L < l_R$. 

An estimation for $l_R$ can be extrapolated through an exponential fit of the region between the two peaks in the distributions of Fig. \ref{fig:skew}, by converting the angular scale into a crystal-depth scale. As an example, for the case of $R_2$, the outcomes of the fits gave $l_{R_2,exp} = (0.19 \pm 0.05) mm$ in agreement with the simulated value $l_{R_2,sim} = (0.20 \pm 0.03) mm$.

\begin{figure}[h!]
		\center{\includegraphics[width=1\columnwidth]{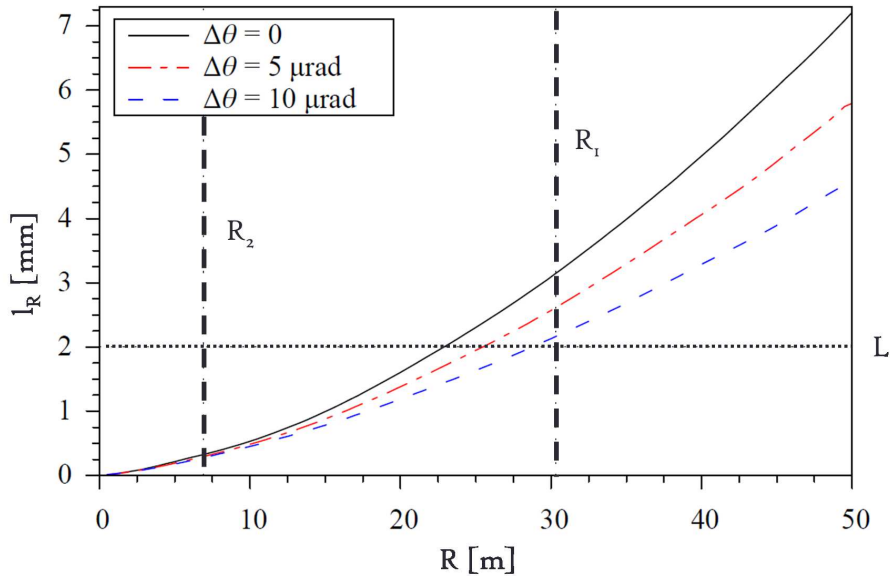}}
		\caption{Dependence of the \textit{relaxation length}, $l_R$, to the curvature radius for three different initial beam divergences; zero (solid line), 5 $\mu$rad  (dashed-dotted line) and 10 $\mu$rad (dashed line). The two vertical lines corresponds to the experimental values $R_1 =$ 30.3 m and $R_2 =$ 6.9 m. The horizontal line represents the experimental crystal length $L =$ 2 mm. The simulations do not take into account the experimental resolution. In the Monte Carlo a particle is considered as \textit{escaped} from SD if the angle between its momentum and the atomic string exceeds the critical angle for axial channeling.}
	\label{fig:relax}
\end{figure}

We carried out a series of simulations to study the dependence of $l_R$ on the bending radius. Fig. \ref{fig:relax} displays the results of the simulations for three different initial beam divergences. The horizontal dot-line highlights the crystal length, while the two dot-dashed vertical lines the radii of curvature used in the experiment. One can notice that as $R=R_1$, $l_R$ slightly exceeded $L$, while at $R = R_2$ the opposite condition, $L \gg l_R$, holds. Therefore, $l_R$ can be used to establish the condition for efficient steering or splitting of positively charged particles as follows:
\begin{description}
  \item[Condition A] If $L \leq l_R$, the crystal behaves as a total \textit{beam steerer} via Stochastic Deflection;
  \item[Condition B] If $L \gg l_R$ and $ L \ll l_D$, the crystal behaves as a \textit{beam splitter}.
\end{description}

One can experimentally exploit the two deflection regimes defined by conditions A-B to effectively accomplish operations that are typical of the accelerator physics. In particular, the usage of a bent crystal respecting condition A as a passive element in a crystal-based collimation scheme offers not only the advantage of reducing the amount of material of the primary collimator, but also accomplishes deflection of the whole beam to the same direction through a single pass, an operation which cannot be done via currently used scheme exploiting PC \cite{Scandale201078}, where multiturn interaction is mandatory. A further advantage of SD, is the reduction of interactions of protons with the nuclei of the crystals as compared to planar case \cite{Chesnokov2014118}, thus reducing the amount of particles in the secondary halo. 

\begin{figure}[h!]
		\center{\includegraphics[width=1\columnwidth]{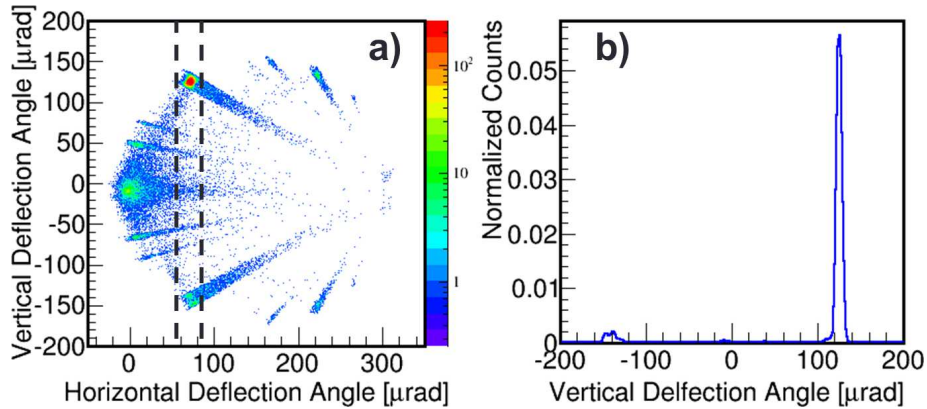}}
		\caption{a) Simulated angular distribution of 400 GeV/c protons after interaction with the 2 mm long Si crystal bent with curvature radii equal to $R_2$ = 6.9 m, with an incidence angle $\theta_{X,in}=-$15 $\mu$rad, $\theta_{Y,in}=$8.66 $\mu$rad with respect to the $\left< 111 \right>$ axes. b) Vertical deflected profile in the two (110) skew planes, obtained via projection of the 2D-distribution of Fig. \ref{fig:fig5}a enclosed between the two vertical dashed-lines.}
	\label{fig:fig5}
\end{figure}

On the other hand, a crystal that meets the condition B can effectively be used for beam splitting. In such a case, a crystal-based extraction can be exploited to set up an extracted beam layout on two experimental channels in just one extraction point from a high-energy hadron accelerator as LHC or FCC. Moreover, this scheme offers the advantage to decrease the radiation damage of the absorbers used for beam collimation (by increasing the number of absorption points and thereby the absorption area). Finally, the axial-to-planar relaxation process offers the possibility to adjust the intensity on the two beams by properly tuning the beam direction with respect to the crystal axis. Indeed, if the incidence angle of the beam with respect to the crystal axis is different from zero and closer to one of the two skew planes, most of the particles preferentially relax only on that plane, as demonstrated by simulation shown in Fig. \ref{fig:fig5}a. Here, most of the particles are captured by the (01$\bar{1}$) skew plane, as highlighted in Fig. \ref{fig:fig5}b, which compares the beam intensity of the two (110) skew planes.

\section{Conclusions}

In summary, an investigation on the mechanism of relaxation of axially confined 400 GeV/c protons to planar channeling in two 2-mm bent Si crystals was carried out at H8 extracted beamline of CERN-SPS. The study of relaxation allowed one to individuate useful conditions (A and B) to exploit a short passive bent crystal oriented with its main axis to the beam as a tool in particle accelerator. In the first mode (A), total relaxation is prevented and the crystal may be used as an efficient deflector for the whole beam. In the second regime (B), relaxation is favored and a short Si crystal can be exploited to efficiently split the beam into two separated beams with adjustable intensity.

\section{Aknowledgments}

We recognized partial support of the INFN-CHANEL experiment, of the IUSS-grant by University of Ferrara, of the the National Academy of Sciences of Ukraine (project no. CO-7-1) and of the Ministry of Education and Science of Ukraine (project no. 0115U000473). Authors acknowledge the help and support of SPS coordinator and staff. We also thank G. Frequest of Fogale Nanotech (Nimes, France) for his support in measuring the crystal thickness and A. Persiani and C. Manfredi of Perman (Loiano, Italy) for their support with crystal holder manufacturing.

\bibliographystyle{unsrt}
\bibliography{biblio}

\end{document}